# Pseudo-Label Guided Multi-Contrast Generalization for Non-Contrast Organ-Aware Segmentation

Ho Hin Lee[1], Yucheng Tang[2], Riqiang Gao[1], Qi Yang[1], Xin Yu[1], Shunxing Bao[1], James G. Terry[3], J. Jeffrey Carr[3], Yuankai Huo[1], Bennett A. Landman[1,2,3]

[1]Department of Computer Science, Vanderbilt University, Nashville, TN, USA 37212;
[2]Department of Electrical and Computer Engineering, Vanderbilt University, Nashville, TN, USA 37212;
[3]Radiology, Vanderbilt University Medical Center, Nashville, TN, USA 37235

**Abstract**: Non-contrast computed tomography (NCCT) is commonly acquired for lung cancer screening, assessment of general abdominal pain or suspected renal stones, trauma evaluation, and many other indications. However, the absence of contrast limits distinguishing organ in-between boundaries. In this paper, we propose a novel unsupervised approach that leverages pairwise contrast-enhanced CT (CECT) context to compute non-contrast segmentation without ground-truth label. Unlike generative adversarial approaches, we compute the pairwise morphological context with CECT to provide teacher guidance instead of generating fake anatomical context. Additionally, we further augment the intensity correlations in "organ-specific" settings and increase the sensitivity to organ-aware boundary. We validate our approach on multi-organ segmentation with paired non-contrast & contrast-enhanced CT scans (n=56) using five-fold cross-validation. Full external validations are performed on an independent non-contrast cohort (n=29) for aorta segmentation. Compared with current abdominal organs segmentation state-of-the-art in fully supervised setting, our proposed pipeline achieves a significantly higher Dice by 3.98% (internal multi-organ annotated, $p < 0.01$), and 8.00% (external aorta annotated, $p < 0.01$) for abdominal organs segmentation. The code and pretrained models are publicly available at https://github.com/MASILab/ContrastMix.

Keywords: Non-contrast CT, unseen domain adaptation, abdominal organ segmentation

## 1. Introduction

Contrast-enhanced CT (CECT) is a routine imaging modality to enhance the conspicuity of tissues of interest [1]. Different contrast phases of imaging are reconstructed with the injection of contrast agents according to the time contrast agents retraining in the blood. For example, CECT can demonstrate a higher intensity contrast for abdominal organs (e.g., aorta, liver, spleen, kidney) to improve the visibility and map tissue permeability of the anatomies (e.g. blood vessels) that could not be typically seen using regular non-contrast CT (NCCT) as shown in Figure 1 [2]. Currently, deep learning methods have demonstrated their capability in segmenting abdominal organs on CECT [3-5]. By contrast, relatively few prior studies have been developed for organ segmentation on non-contrast CT scans, especially those whose boundaries are difficult to distinguish from the nearby tissues (e.g. aorta). However, the abdominal NCCT is more widely and conveniently available than compared with the CECT, and has been broadly used. For example, NCCT is routinely acquired as a diagnostic modality for detecting renal stone or intramural hematoma in the aorta [6, 7]. Therefore, it is appealing to develop deep learning-based segmentation methods that can achieve precise tissue segmentation on NCCT in the absence of contrast-enhanced boundaries between neighboring organs. Unfortunately, deep learning models that are trained with CECT only yield degraded segmentation performance when directly applied to NCCT scans. This phenomenon is well-known as the limited generalizability of transferring deep learning methods across different domains [8, 9].

In the natural imaging domain, the domain shift may contribute to different factors (e.g. illumination, pose and image quality). Such data distribution shifts lead to the degradation of performance. Previous works demonstrate utilizing the labeled data is needed in one or more relevant source domains to execute the new task in the target domain. However, the distribution shift is comparatively lower than the different modalities in medical imaging. The domain shift can be further classified into two main aspects in the medical imaging domain: contrast intensity and shape morphology [10, 11]. With different protocols or phases, significant intensity variations may exist across additional CT scans. To tackle the challenges from domain shift, previous works have been designed in two main directions: 1) unsupervised domain adaptation (UDA) [12] and 2) multi-source domain generalization (MDG) approaches [13]. Adversarial generative approaches are proposed to perform image translation from the source domain to the target domain [14, 15]. However, imbalanced training data exists between domains and the style transfer may lead to a significant loss of anatomical context in each organ of interest. The generative approaches are limited to generate a similar style with refined anatomical details of organs with stability. Moreover, a generative model is often unavailable. Thus, it is difficult to deploy the generative model in clinical practice. For MDG approaches, domain-invariant features such as shape topology are aimed to extract and apply the representation to the unseen domain dataset [16]. However, when the anatomical context is not aligned in a well-defined spatial reference, it is difficult to adapt the correspondence between the organ topology across domains. With the above limitations to the current state-of-the-art, we aim to 1) stably adapt the cross-domain dataset that minimizes the loss of anatomical context and generalizes the contrast variation across domains; and 2) extract the dense correspondence between the organ topology as the domain-invariant representation for learning.

In this work, we present a novel 3D anatomical-aware learning scheme ContrastMix, to perform robust organ segmentation on NCCT scans (target domain) with CECT scans (source domain) as teacher guidance. The backbone of the segmentation method is inspired by RAP-Net [4], which provides high flexibility to promote robust 3D abdominal multi-organ segmentation using single model architecture. The learning scheme uses the pairwise samples that incorporate significant shifts in contrast intensity and morphology. Specifically, we apply an intra-subject registration to efficiently align NCCT to CECT scans to minimize the morphological shift. Meanwhile, volumetric patches are extracted for each organ of interest from the coarse volumetric segmentation from source domain as a spatial prior. We generate a refine probability mapping with source domain patches using a teacher network and use it as soft supervision. Moreover, we sharpen the self-predicted context of non-contrast patches using knowledge distillation during the training process and decrease the uncertainty of the boundary region. Furthermore, a weighting parameter is sampled randomly from a beta distribution, which mix both the contrast (image) and morphological (segmentation prior) variation from both domains to adapt the subtle boundary information with non-contrast characteristics. The experimental results demonstrated that proposed ContrastMix outperforms the state-of-the-art supervised baselines with consistent improvements across internal and external testing cohorts. Our contributions are four-fold:

- We propose an unsupervised learning approach that adapts the pairwise contrast-enhanced domain knowledge and self-predicted non-contrast domain as pseudo supervision for refining organ segmentation.

- We propose a knowledge distillation technique to sharpen the boundary context and use teachers refined context in the contrast-enhanced domain as soft supervision to preserve the morphology of organs.
- We propose a local mixing technique that efficiently adapts the diverse contrast appearance and enriches the boundary information of each organ across domains.
- We validated our approach using two separate non-contrast clinical cohorts with various imaging scanners and protocols, and demonstrated a significant improvement comparing with current hierarchical and transformer-based state-of-the-art approaches.

2. Related Work

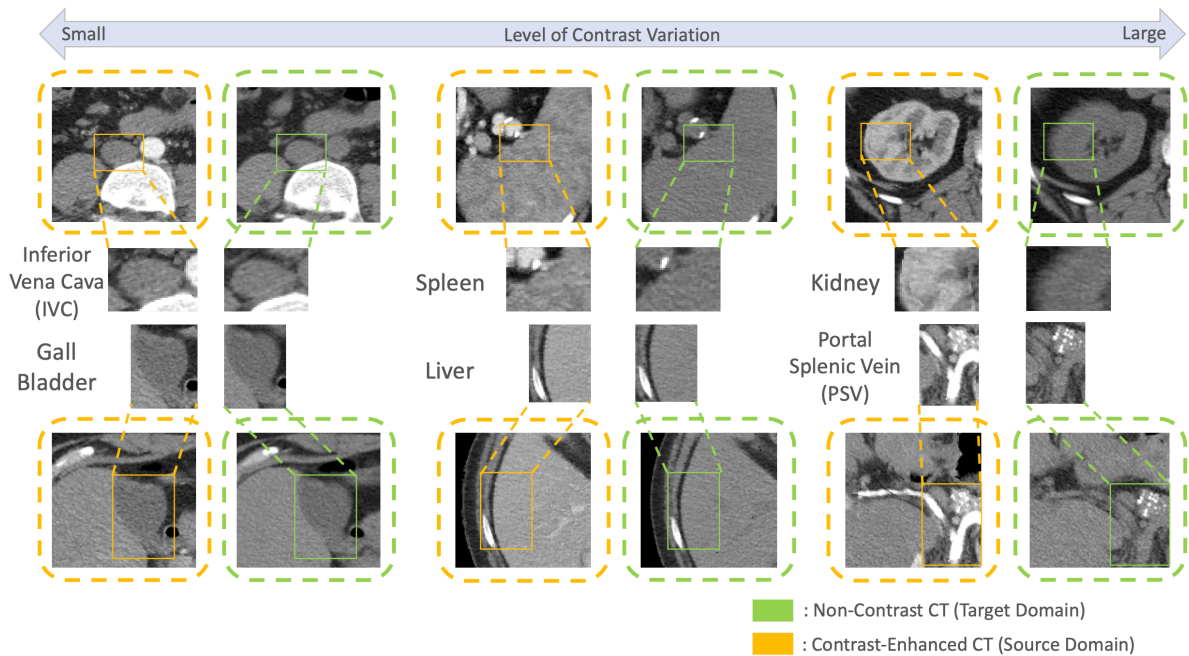

Figure 1. In multi-contrast phase CT, the contrast intensity varies according to the time of contrast agent retaining in the blood vessels. Different levels of contrast variation are demonstrated across the organ interest. The robustness of model that pre-trained with contrast-enhanced CT is limited to adapt the anatomical context across all organs in non-contrast CT. It is challenging to identify subtle boundary information between neighboring organs without additional contrast guidance.

**Medical Image Segmentation:** Previous efforts have tackled organ segmentation in contrast-enhanced CT with significant usage of deep learning. A naive approach of aggregating target domain data is to directly train a deep network model with ground-truth label and has been validated with various regions of CT in the existing literature [17, 18]. However, the supervised strategies require a large a number of high-quality ground truth labels and the inaccuracy of the boundary information still exists between neighboring anatomical structures. Meanwhile, volumetric images have to be downsampled to fit into deep neural networks for training due to the limited GPU memory [19, 20]. To minimize the loss of anatomical context, patch-based approaches are proposed to adapt the high-resolution context for segmentation [21] . *Huo et al.* proposed patch-based methods for whole-brain segmentation [22]. However, the patch-based approaches only adapt the local representation and are limited in adapting the global spatial information in complete images. Therefore, hierarchical systems are proposed to adapt feature representation across scales. *Roth et al.* proposed a coarse-to-fine method that roughly defines the local region to extract representation for refined segmentation [23]. *Roth et al.* further extended the coarse-to-fine process into a multi-scale pyramid network to perform segmentation in high resolution [20]. However, the performance is limited due to the inaccuracy prediction on the bounding box with the low-level models. In addition, upsampling is needed to resample the representation back to the original image resolution. *Li et al.* demonstrates a hybrid algorithm to integrate the 2D slice representation from the first stage with the 3D volumetric context in the second stage [24]. Apart from adapting 2D information with 3D representation, *Zhu et al.* proposed an effective sliding window approach to extract the region of interest for refining segmentation [25]. Additionally, an expanding bounding box is used to allocate the target regions accurately and minimize the outliers (out of target regions) for training models. To further integrate the global context for refining local segmentation, *Tang et al.* proposed a high-resolution approach to adapt the target-corresponding coarse segmentation as the additional channel with volumetric image patches for model training [5]. However, *Tang et al.* and *Zhu et al.* targeted on the single organ segmentation in the refine stage and are limited to adapt the morphological variability across the organs in image patches. *Lee et al.* proposed RAP-Net to use the pseudo-binary organ prior as additional guidance and adapt all organ patches in a single network architecture [4].

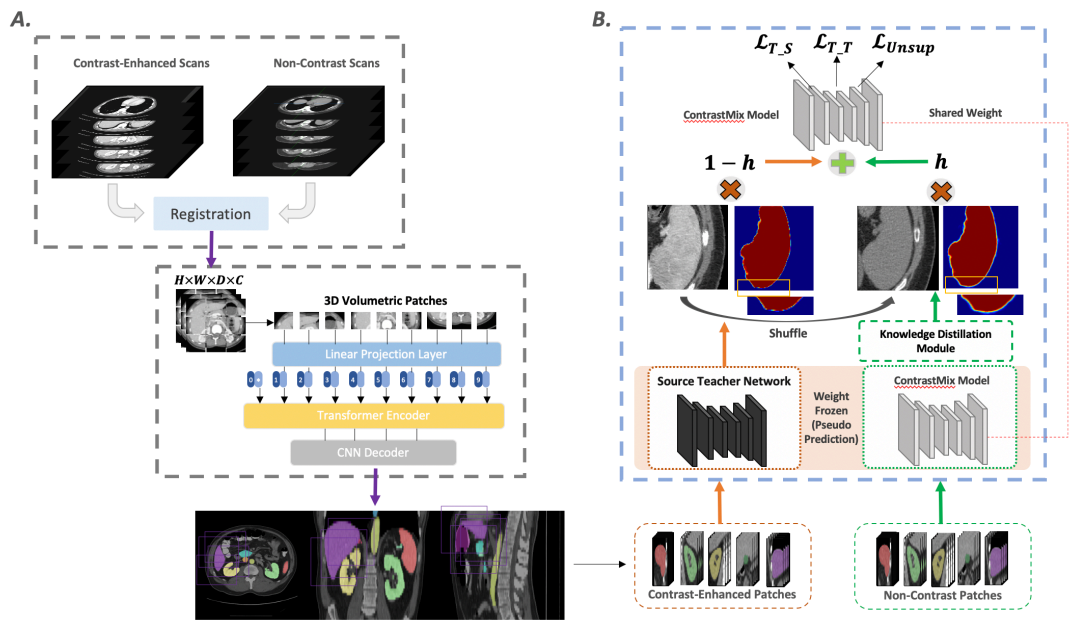

Figure 2. Summary of the proposed method: **A. Minimize anatomical shift across domains:** Intra-subject registration is initially performed to minimize the anatomical change of organ interests between domains. A hierarchical registration approach with 1) DEEDS affine registration and 2) DEEDS deformable registration is used to transfer the non-contrast context towards contrast-enhanced spatial reference. Coarse segmentation is then generated for both domains imaging with a pre-trained transformer model to adapt the localization information of each organ interest. **B. Generalize the contrast variation across domains:** Corresponding organ patches are extracted and the coarse segmentation prior is concatenated with the image as additional channel to perverse the global context. Refine prior knowledge is generated with a teacher model for source domain, while knowledge distillation is performed to sharpen the probability of the boundary region for the self-predicted target domain representation. A random weighting parameter is computed to generate different contrast intensity combinations for images patches and enrich the boundary context by combining the weighted probability mapping from each domain.

**Domain Adaptation for Segmentation:** On top of deep supervised training, several studies performed domain adaptation techniques to find additional context for supervision from cross-modality data and compute segmentation on the target domain. With the investigation of generative adversarial approaches, several works proposed an image translation network (GAN-based) to align the contrast appearance from source to target domain and use the generated image for further processing. *Huo et al.* adapts the domain invariant feature across MRI to CT by using adversarial neural networks and further adds a segmentation module to increase the stability of the adaptation framework [14]. *Dou et al.* investigates the extraction of domain-invariant features in feature space [26]. In contrast, *Chen et al.* demonstrates a synergistic method to adapt domain-invariant features in both feature and image space [15]. *Chandrashekar et al.* synthesizes images with contrast from the non-contrast domain and adjusts the contrast-characteristics for segmentation [27]. However, the dataset from different domains may be unbalanced. It is challenging to translate the contrast characteristic of each organ interest into the target domain with stability. Apart from the generative approach, the multi-modal approach is used to leverage the modality-shared knowledge using feature fusion strategies. *Valindia et al.* [28] and *Tulder et al.* [29] demonstrate different parameter sharing strategies for unpaired multi-organ segmentation, while *Dou et al.* introduce the modality-specific normalization to adapt cross-modality unpaired context using knowledge distillation [30]. Although promising progress is demonstrated, the adaptation framework is limited in 2D networks with a high demand for sufficient ground truth labels for volumetric image segmentation.

## 3. Method

An overview of the proposed method is demonstrated in Fig. 2 and our goal is to perform robust abdominal organs segmentation for non-contrast CT in a unsupervised setting. Let $S = \{(x_s, y_s)\}_{s=1}^{|S|}$ be the set of source domain contrast-enhanced cohorts and $T = \{(x_t, y_t)\}_{t=1}^{|T|}$ be the set of target domain non-contrast cohort, where each $x_s, x_t \in \mathbb{R}^{|\Phi|}$ is the corresponding domain paired image and $y_s, y_t \in \{0,1\}^{|\Phi| \times |C|}$ is the corresponding one-hot paired ground truth segmentation mask. Here, we denote $\Phi \in \mathbb{R}^2 \times \mathbb{Z}$ as the dimension of images voxels and $C \in \mathbb{R}$ as the number channels for the segmentation classes. The proposed pipeline consists of three main hierarchical components: (1) intra-modal registration for generating anatomical context, (2) knowledge distillation of target domain self-supervised context and (3) cross-domain intensity-prior mixing with random weighting.

### 3.1 Intra-Modal Registration for Anatomical Prior Generation

As the organ anatomies are not well aligned in the initial state of both target and source domains, intra-modal registration is performed to minimize the anatomical shift between domains in an intra-subject setting. Here, we apply DEEDS (DEnsE Displacement Sampling) to perform hierarchical two-stage registration [31], which consists of 1) DEEDS affine registration and 2) DEEDS deformable registration (Fixed image: non-contrast domain, moving image: contrast-enhanced domain). The contrast correspondence between organs can be well demonstrated and extracted pairwise anatomical context as voxel-to-voxel. A transformer-based approach generates coarse segmentation with complete volumetric inputs from source domains [32]. The volumetric images are downsampled to a specific resolution with tri-linear interpolation. Consider the coarse segmentation network $F_c$ is parameterized with $\theta_c$, the segmentation model aims to minimize as following:

$$\arg\min_{\theta_c, y} \mathcal{L}_V(\theta_c)$$

Here, we denote $\mathcal{L}_V$ as a combination of soft Dice loss and cross-entropy loss to train low-resolution segmentation model with multiple semantic targets. The corresponding definition is as follows:

$$\mathcal{L}_V = 1 - \frac{2}{A}\sum_{a=1}^{A}\frac{\sum_{i=i}^{I} y_{i,j} h_{i,j}}{\sum_{i=1}^{I} y_{i,j}^2 + \sum_{i=1}^{I} h_{i,j}^2} - \frac{1}{I}\sum_{i=1}^{I}\sum_{j=1}^{J} y_{i,j} log(h_{i,j})$$

where $A$ denotes as the number of anatomical class; $I$ is the number of voxels and $h_{i,j}$ is the probability output with softmax activation for class $A$ at voxel $i$. The one-hot localization $a_s, a_t \in \{0, 1\}^{|\Phi| \times |C|}$ provides prior knowledge to adapt organ-wise settings (organ-corresponding patches) for refined segmentation. With the intra-modal registration, images from both domains are anatomically aligned. Therefore, we randomly select voxels within each organ's one-hot bounded regions, and multiple bounding boxes are placed as local views to extract volumetric patches with the voxels chosen as a center point. The extracted image patches are concatenated with the prior context as the additional channel guidance and perform organs segmentation refinement end-to-end in a single network architecture.

### 3.2 Knowledge Distillation of Self-Supervised Target Domain Context

Due to the non-contrast characteristic, it is challenging to identify subtle boundaries and may lead to over-segmentation across neighboring organs. To tackle such limitations, we propose to sharpen the boundary information and increase the probabilistic certainty of the self-predicted pseudo segmentation. We initially performed $K$ augmentation for both images and coarse prior $\hat{x}_{t,k}, \hat{x}_{s,k}, \hat{a}_{t,k}, \hat{a}_{s,k} = Augment(x_t, x_s, a_s, a_t), k \in (1, \dots, K)$. Both augmented image and coarse prior are then concatenated as multi-channel input $m_{t,k}, m_{s,k}$ to preserve the representation learned within the localized regions. For contrast-enhanced patches, a teacher model $q_s$ is used to compute the probability map and provide sufficient boundary context to refine target segmentation. For non-contrast patches, the student segmentation network $F$ is trained from scratch and generates self-predicted probability maps $q_t$ as self-supervision. We denote $N$ as the number of the data augmentations and compute the average prediction $\tilde{q}_t$ across all self-predicted probability maps within a minibatch via softmax activation if $N > 1$:

$$\tilde{q}_t = \frac{1}{K}\sum_{k=1}^{K} F(m_{t,k}; \theta)$$

As the average prediction may extract inaccurate boundary information from the self-predicted context, we introduce a knowledge distillation module to increase the sensitivity towards the target organs' boundary. The knowledge distillation function is defined as follows:

$$S(\tilde{q}_t, T)_i = \tilde{q}_{t_i}^{\frac{1}{T}} \Big/ \sum_{j=1}^{C} \tilde{q}_{t_j}^{\frac{1}{T}}$$

where $\tilde{q}_t$ is the average source domain probability map with $C$ classes prediction over $K$ augmentations with knowledge distillation, and $T$ is a temperature scalar to sharpen the soft prediction labels and amplify the inter-class relationships in the self-prediction stage.

### 3.3 Cross-Domain Intensity-Prior Mixing with Random Weighting

Apart from the limitation of subtle boundary, the contrast variation between organ interests poses a great challenge to the segmentation generalizability across domains. Here, we propose an intensity-prior mixing module to random weight the contrast intensity and prior context of each domain, and sum it to generate an augmented version as the generalized representation. Specifically, we first concatenate both domains input and generate a shuffle version of the concatenated input $(\tilde{m}_1, \tilde{m}_2)$. Both concatenated images and pseudo probability mapping are multiplied with a random weighting $h$, which is sampled from a beta distribution $h \sim Beta(\alpha, \beta)$ with hyperparameters $\alpha$ and $\beta$. The mix-and-match strategy provides another form of data augmentation for the segmentation model $F$ to behave linearly in-between the cross-domain samples. Let $p = F(x; \theta) \in \{0,1\}^{|\Phi| \times |C|}$ be the class probability map prediction from the segmentation network. Both the contextual features extracted from both source and target domains are integrated as $(x', p')$ by:

$$x' = h \cdot (m_s, m_t) + (1 - h) \cdot (\tilde{m}_1, \tilde{m}_2)$$

$$p' = h \cdot (p_s, p_t) + (1 - h) \cdot (\tilde{p}_{m_1}, \tilde{p}_{m_2})$$

The context shuffling strategies increase the confidence level of predicting uncertain regions between the neighboring organs with the integration of source domain representation. Additionally, we use $p'$ as soft guidance to preserve the topology of organ interests and benefit to identify the subtle boundary with non-contrast characteristics. The weighted samples are then input into the segmentation network and computed to refine binary segmentation as the final output.

### 3.4 Loss Functions

Assume that the non-contrast segmentation network $F$ is parameterized by the set of weights $\theta$, both target and source domain patches are exploited jointly in the training process by minimizing the following loss function:

$$\mathcal{L}_{total} = \mathcal{L}_{T\_s}(\theta; S) + \mathcal{L}_{T\_t}(\theta; T) + \lambda \mathcal{L}_{unsup}(\theta; T)$$

To preserve the similarity of the organ anatomy between pairwise images, teacher knowledge supervision loss $\mathcal{L}_{T\_s}(\cdot)$ and $\mathcal{L}_{T\_t}(\cdot)$ are computed to constrain the segmentation network and generate predictions with a similar anatomical structure to the teacher model predictions. Besides optimizing the non-contrast target outputs, constraining the contrast-enhanced prediction can help adapt the shape invariant representation in the source domain imaging. The teacher knowledge supervision loss is defined as follows:

$$\mathcal{L}_{T\_s}(\theta; S) = \sum_{(x_s, y_s) \in S} \ell_{T_s}(F(x_s; \theta), M_s)$$

$$\mathcal{L}_{T\_t}(\theta; T) = \sum_{(x_t, y_t) \in T} \ell_{T_t}(F(x_t; \theta), M_s)$$

We denote $M_s$ as the teacher model prediction with source domain images. Both $x_s$ and $x_t$ are registered aligned in the preprocessing step and well adapt to the topological correspondence across organs in the abdominal region. Dice loss is used for the segmentation loss functions $\ell_{T_s}(\cdot)$ and $\ell_{T_t}(\cdot)$. The dice loss function is defined as follows:

$$\ell_{T_t}(p_t, M_s) = \sum_{p_t \in T} \left(1 - 2 \cdot \frac{p_t \cdot M_s}{p_t + M_s}\right)$$

$$\ell_{T_s}(p_s, M_s) = \sum_{p_s \in S} \left(1 - 2 \cdot \frac{p_s \cdot M_s}{p_s + M_s}\right)$$

Apart from the teacher supervised loss, a soft unsupervised loss function is employed to further leverage the self-predicted context and minimize the probability of over-segmentation. We introduce $\mathcal{L}_{unsup}$ to combine the cross-entropy loss between $p_s$ and the teacher prediction $M_s$, and the square $L_2$ loss on the final target domain prediction and the self-supervised intermediate prediction $M_t$. The integration with square $L_2$ loss reduces the sensitivity of incorrect prediction within the anatomical context bounded region. The unsupervised loss is defined as follows:

$$\mathcal{L}_{unsup} = -\sum_{i \in \Phi} \sum_{j \in C} M_{s_{ij}} \log\left(p_{s_{ij}}\right) + \lambda_t \sum_{p_t \in T} ||p_t - M_t||_2^2$$

where $\lambda_t$ is the weighting hyperparameter of the unsupervised loss computed with the self-predicted outputs. Overall, the training objectives including $\mathcal{L}_{T\_s}$, $\mathcal{L}_{T\_t}$ and $\mathcal{L}_{unsup}$ are optimized end-to-end concerning weighting parameters $\theta$. In the testing phase, only non-contrast target domain scans are input for inference and obtain refined segmentation with a majority vote for joint label fusion.

## 4. Experiments

To evaluate the performance of the proposed multi-contrast domain adaptation pipeline, we test it on four medical image segmentation tasks involving contrast-enhanced CT and non-contrast phase CT samples. In the following subsection, we describe more details about the datasets that we implement, implementation details of our pipeline, experimental setup, and the evaluation metrics for both validation and testing in internal and external cohorts (U54DK120058).

### 4.1 Datasets

We have conducted experiments on two independent clinical cohorts, consisting of 1) Clinical research cohorts with contrast-enhanced and non-contrast pairwise CT volumes for multi-organ segmentation tasks and 2) healthy clinical cohorts with non-contrast CT volumes for aorta segmentation tasks. All three datasets are retrieved in de-identified form from Vanderbilt University Medical Center (VUMC) under IRB (Institutional Review Board) approval.

**Contrast-Enhanced & Non-Contrast Pairwise Clinical Research Cohort (CENC)**: We retrieved 56 de-identified splenomegaly subjects with pairwise portal venous phase CT and non-contrast phase CT for internal training and validation. Each contrast-enhanced and non-contrast volume is annotated and refined manually with 12 classes of multiple abdominal organs. The ground truth labels of 12 multiple organs are provided including 1) spleen, 2) right kidney, 3) left kidney, 4) gall bladder, 5) esophagus, 6) liver, 7) stomach, 8) aorta, 9) inferior vena cava (IVC), 10) portal splenic vein (PSV), 11) pancreas and 12) right adrenal gland. The axial-plane pixel dimension of each scan varies from 0.64 to 0.98 mm and the corresponding slice thickness (z-axis) is constantly 3mm. Each CT volume consist 75 -116 slices with 512 × 512 pixels.

**Non-Contrast Healthy Clinical Research Cohort (NCH)**: We retrieved 29 de-identified subjects free from splenomegaly with non-contrast phase CT for external validation of non-contrast segmentation performance. Manual annotation of aorta volumes (the ground truth) was performed by expert image analysts under the supervision of a clinical radiologist (MD) from Vanderbilt University Medical Center. These scans have a large variance of the morphology for the aorta, with volumes varying from 39.3 cubic centimeters (cc) to 96.9 cc. Each CT volume consists of 70 slices of 512 × 512 pixels and has a constant resolution of 0.68 mm × 0.68 mm × 3.00 mm for axial plane and slice thickness across all subjects.

### 4.2 Implementation Details

#### 4.2.1 Preprocessing

Both contrast-enhanced phase and non-contrast phase CTs are initially processed with soft tissue window and voxel intensity range limit from -175 to 250 Housefield Unit (HU). As clinically acquired CT scans may be obtained with complete body or limited field of view for the abdominal region, we perform a body part regression network to predict the approximate score representing the anatomical location on the body [33]. We limit the score to a range from -4 to 5 corresponding to the abdominal region and crop the volumes to localize the abdominal region. Intensity normalization is performed for each CT volume after applying soft tissue window with min-max normalization. The voxel intensity value is normalized with a range from 0 to 1 for training the segmentation network.

#### 4.2.2 Implementation Details

We first implemented a transformer network UNETR [32] as the volume-based segmentation model and further employed a 3D U-Net [19]like architecture network as the patch-based model hierarchically. For the transformer network, all images are initially downsampled to a resolution of $1.5 \times 1.5 \times 1.5$ and a 1D sequence is generated by dividing the samples into flattened uniform non-overlapping patches with a dimension of $96 \times 96 \times 96$. Twelve transformer blocks are used, comprising of multi-head self-attention modules and multilayer perceptron sublayers. The sequence representations in transformer blocks 3, 6, 9, and 12 are extracted and concatenated with the upsampled feature from each decoder block with a factor of 2. The final layer output is fed into a $1 \times 1 \times 1$ convolutional layer with 13 output channels and generates semantic segmentation with a softmax activation function.

For the cross-domain refine segmentation, the segmentation model consists of 8 encoders with a convolutional kernel size of $3 \times 3 \times 3$, batch normalization layers, and 10 decoders with the deconvolutional size of $2 \times 2 \times 2$. Skip connections between encoder and decoder are used to integrate and capture the variant representations from encoder blocks across scales. A $1 \times 1 \times 1$ convolutional layer is used as the final layer to output the number class label for segmentations. We randomly pick 50 voxel points for each organ in the volumetric anatomical prior from the coarse model. The chosen voxel points are localized as the center point and extract patches with a dimension of $128 \times 128 \times 48$. In total of 600 volumetric patches are extracted for all 12 organs each volume for training. To increase the reproducibility and perform fair comparison, both volume-based and patch-based networks, use the same hyper-parameters except the dimension and channel number of inputs. We perform training with a batch size of 1 and adapt ADAM Weighted algorithm with stochastic gradient descent (SGD), momentum=0.9 as the optimizer for both networks. The learning rate is 0.0001 and reduces by a factor of 0.9 every 5 epochs. All experiments are implemented on NVIDIA 2080 Ti 11 GB GPU with CUDA 10.0. The code of all experiments including baseline methods and ablation studies are implemented in python 3.8 with Pytorch version of 1.9.0.

### 4.2.3 Evaluation Metric

We employed two common metrics for evaluating the segmentation performance of all testing methods with ground truth labels: 1) Dice similarity coefficient (Dice) and 2) Mean Surface distance (MSD). The definition of the Dice coefficient is to measure the overlapping region between the volumetric prediction of the segmentation model and the ground-truth segmentation label. It is defined as:

$$Dice(P, G) = \frac{2|P \cap G|}{|P| + |G|}$$

where $P$ denotes as the segmentation pipeline prediction and $G$ corresponds to the ground-truth labels. As the Dice value range from 0 to 1, a high Dice score represents good quality of segmentation prediction.

Apart from measuring the volumetric overlapping region, we extract the 3-dimensional coordinates of vertices to measure the mean distance between every point in $P$ and the nearest point in $G$. The mean surface distance metric is defined as:

$$MSD(P, G) = \frac{1}{2}(\bar{d}(P, G) + \bar{d}(G, P))$$

Where $\bar{d}(P, G)$ is the mean of distances between each voxel in $P$ and the nearest surface voxel in G. Two values are computed for both end-diastolic and end-systolic instances. We finally computed the average value of these two instances and the low mean surface distance that corresponds to good segmentation performance.

### 4.3 Experimental Setup

We conducted experiments on two perspectives of analysis to evaluate the effectiveness of the cross-domain adaptation pipeline on non-contrast organs segmentation. We performed multi-organ segmentation on the non-contrast samples in CENC dataset as internal training and validation. The network is trained from scratch with five-fold cross-validations using CENC dataset: training subjects: n=44; validation subjects: n=6; testing subjects: n=6 (10% ratio of the samples are used for testing).

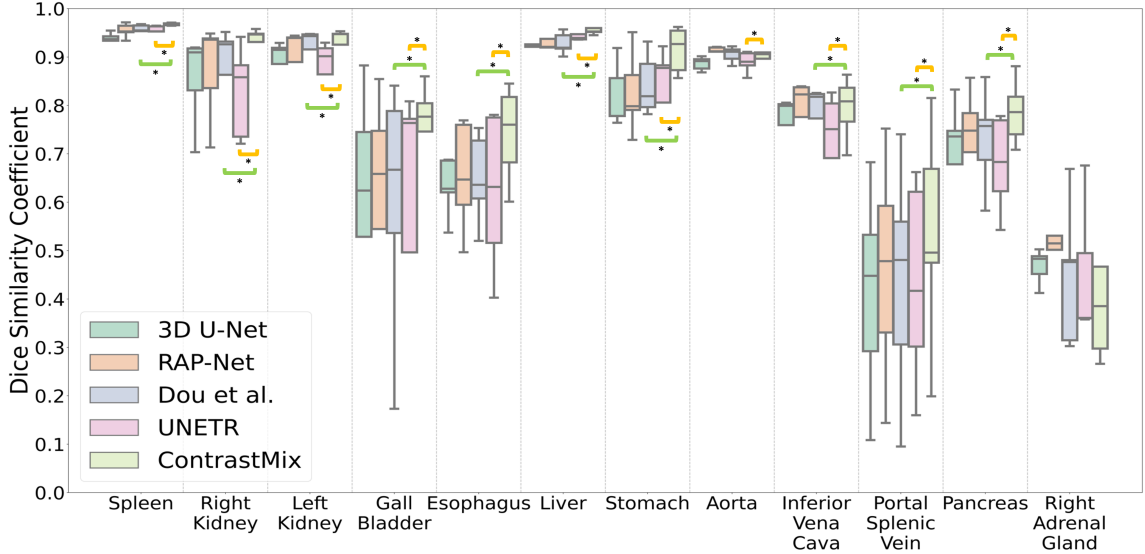

Figure 3: ContrastMix outperforms the current state-of-the-art hierarchical pipeline, multi-modal segmentation approach, and the single-stage transformer-based approach across most organs in the internal dataset CENC inference.

Table 1. Comparison of current state-of-the art with supervised approaches, adversarial generative approach and multi-modal approach on multi-organ segmentation with CENC dataset. (*: p<0.01, with Wilcoxon signed-rank test)

| Method | Spleen | R.Kid | L.Kid | Gall | Eso | Liver | Aorta | IVC | Ave. Dice |
|---|---|---|---|---|---|---|---|---|---|
| 3D U-Net (2016) | 0.937 | 0.856 | 0.912 | 0.690 | 0.631 | 0.920 | 0.880 | 0.769 | 0.762 |
| *Roth et al. (2017)* | 0.940 | 0.870 | 0.923 | 0.701 | 0.674 | 0.925 | 0.891 | 0.772 | 0.770 |
| *Syn-Seg Net (2018)* | 0.941 | 0.868 | 0.910 | 0.654 | 0.652 | 0.927 | 0.895 | 0.780 | 0.764 |
| *Zhu et al. (2019)* | 0.950 | 0.880 | 0.918 | 0.710 | 0.643 | 0.932 | 0.890 | 0.802 | 0.781 |
| *Dou et al. (2020)* | 0.957 | 0.878 | 0.938 | 0.708 | 0.649 | 0.930 | 0.900 | 0.784 | 0.788 |
| *RAP-Net (2021)* | 0.954 | 0.874 | 0.928 | 0.701 | 0.653 | 0.928 | 0.897 | 0.790 | 0.784 |
| *UNETR (2021)* | 0.960 | 0.896 | 0.936 | 0.800 | 0.680 | 0.949 | 0.888 | 0.789 | 0.798 |
| ***ContrastMix*** | **0.971** | **0.926** | **0.950** | **0.820** | **0.736** | **0.960** | **0.915** | **0.821** | **0.820*** |

Moreover, we further evaluate the impact of the distillation module and the cross-domain mixing module on the segmentation performance. We performed ablation studies to analyze the correspondence of the hyperparameter $T$ and $(\alpha, \beta)$ towards the robustness of the segmentation pipeline. Apart from the internal validation, we performed external testing with single organ aorta segmentation within NCH samples using the well-trained model to demonstrate the confidence level on the generalization ability to adapt non-contrast segmentation in an unsupervised setting.

We compared our method against several volumetric supervised state-of-the-arts (SOTA). The fully-supervision SOTA considers all training samples with ground-truth labeled. Our pipeline consists of training the coarse volumetric segmentation network with source domain ground-truth labels; we consider our proposed method an unsupervised setting without using non-contrast ground-truth label usage. Moreover, we compare our pipeline with fully supervised and multi-modal state-of-the-arts methods: 3D U-Net segmentation approach [19], Syn-Seg Net [14], multi-scale segmentation [34], pyramid scaling segmentation [25], RAP-Net [4], segmentation with knowledge distillation [30], and transformer network for segmentation [32]. We constrain all testing inference with the same underlying network architecture, optimization procedure, and data augmentations for a fair comparison. We finally compared the average Dice score and average mean surface distance across organs segmentation and computed the statistical significance with Wilcoxon signed-rank test.

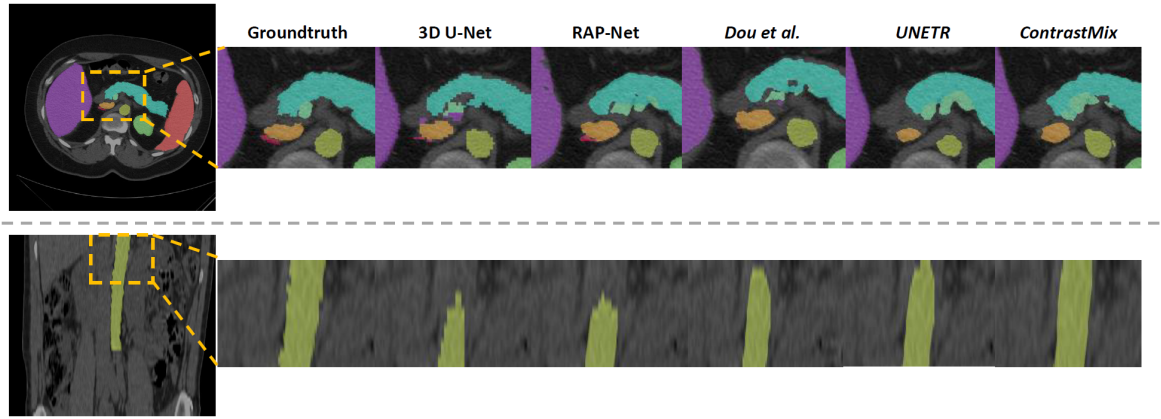

Figure 4: Visualization of segmentations with different state-of-the-art strategies yields incremental improvement in segmentation performance. ContrastMix results in smooth boundaries and accurate morphological information for each organ of interest in both internal and external inferences.

Table 2. Comparison of hierarchical based, multi-modal based and transformer-based state-of-the-art approaches for single organ segmentation on external testing dataset NCH. (*: $p<0.01$, with Wilcoxon signed-rank test)

| Method | Ave. Dice | Ave. MSD |
|---|---|---|
| 3D U-Net (2016) | $0.712 \pm 0.111$ | $4.14 \pm 3.94$ |
| *Roth et al. (2017)* | $0.735 \pm 0.100$ | $3.43 \pm 3.45$ |
| *Syn-Seg Net (2018)* | $0.721 \pm 0.124$ | $3.89 \pm 2.57$ |
| *Zhu et al. (2019)* | $0.746 \pm 0.0978$ | $3.01 \pm 2.21$ |
| *Dou et al. (2020)* | $0.770 \pm 0.114$ | $2.34 \pm 1.88$ |
| *RAP-Net (2021)* | $0.755 \pm 0.0944$ | $2.67 \pm 1.74$ |
| *UNETR (2021)* | $0.775 \pm 0.0820$ | $2.41 \pm 1.84$ |
| ***ContrastMix (Ours)*** | **$0.837 \pm 0.0662$*** | **$1.60 \pm 1.08$*** |

## 5. Results

Figure 3 demonstrated the quantitative comparison of all fully supervised methods in Dice score, including abdomen segmentation SOTA and knowledge distillation segmentation SOTA with the jointly training using CENC dataset. For external validation, Figure 3 demonstrated the quantitative comparison of all state-of-the-arts including fully supervised, semi-supervised, and unsupervised SOTA for the NCH dataset with Dice score and mean surface distance evaluation metric. In additional to qualify the quantitative measurements, Figure 4 further provides an additional qualitative representation of non-contrast samples with inference output overlay and demonstrates the significant improvement in the perspective of the segmentation mask.

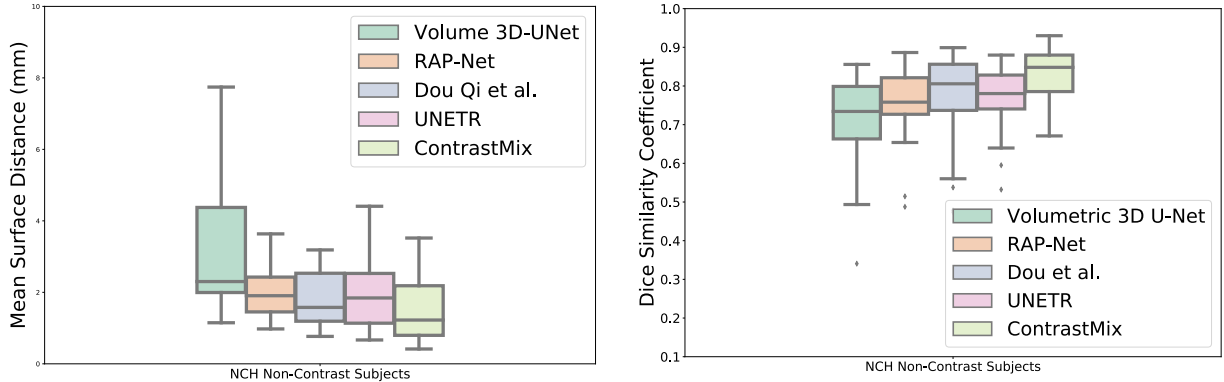

Figure 5: ContrastMix outperforms the current state-of-the-art approaches and demonstrates significant improvement in both Dice score and mean surface distance in the aorta segmentation as external dataset NCH inference.

## 5.1 Internal Testing Performance

As shown in Figure 3 and Table 1, the performance increases consistently from a single U-Net architecture network to a coarse-to-fine approach adapting the non-contrast scans. With the knowledge distillation guidance [30], performance in particular organs such as the spleen and left kidney, improves by reducing the possibility of over-segmentation across organ boundaries. However, Figure 3 illustrates that *Dou et al.* predicts the organ boundaries with less confidence and limits to preserve the core anatomical context of the organs only. UNETR outperforms all the supervised state-of-the-arts with an average Dice score of 0.798 across 12 organs. It further increases the encoder's power to abstract the non-contrast representation context across organs. With the use of ContrastMix, the performance of the non-contrast segmentations significantly boosts from the average Dice score of 0.798 to 0.820 across all organs. The performance on all organs segmentation outperforms significantly to all state-of-the-art methods with statistical significance. The qualitative representation in Figure 4 further shows that ContrastMix refines organ details and preserves the boundary information between neighboring organs.

## 5.2 External Testing Performance

Figure 5 presents the quantitative performance of single organ aorta segmentation with NCH. The trending of the segmentation performance is similar to that of multi-organ segmentation with CENC. Interestingly, the knowledge distillation pipeline significant improves the performance of the state-of-the-art hierarchical RAP-Net and achieves the minimal MSD across all state-of-the-arts. In addition to the transformer network, UNETR shows improvement to a small extent in Dice score and limits to establish an additional advantage in MSD. By using ContrastMix, the segmentation performance outperforms all state-of-the-art approaches with 8.00% leverage in Dice score and 33.6% decrease in MSD. In the qualitative perspective in Figure 4, the segmentation mask of the aorta organ is incrementally refined and ContrastMix best preserves the anatomical and boundary details across all the state-of-the-art approaches.

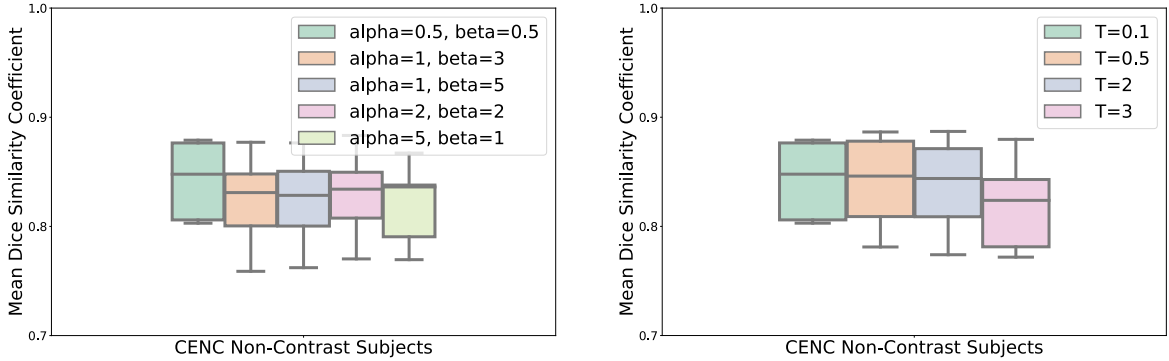

Figure 6: The ablation studies on the contribution of adaptation modules with hyperparameters a) alpha & beta for beta-distribution and b) temperature. From 6(a), we found that the weighting parameters of beta-distribution (alpha = 0.5, beta = 0.5) generate a better performance than that of other beta-distribution. Also, sharpening the self-predicted pseudo labels with T < 1 can help increase the stability on segmentation with 6(b).

Table 3. Ablation studies of multi-organ segmentation performance with the variability of parameters in knowledge distillation and the random weighting parameter generated from different beta-distribution for contrast mixing.

| ContrastMix | Spleen | R.Kid | L.Kid | Gall | Eso | Liver | Aorta | IVC | Ave. Dice |
|---|---|---|---|---|---|---|---|---|---|
| $T = 0.1$ | **0.971** | **0.926** | **0.950** | **0.820** | 0.736 | **0.960** | **0.915** | **0.821** | **0.820** |
| $T = 0.5$ | 0.968 | 0.925 | 0.944 | 0.805 | 0.713 | 0.958 | 0.907 | 0.797 | 0.814 |
| $T = 2$ | 0.969 | 0.921 | 0.945 | 0.790 | 0.716 | 0.957 | 0.909 | 0.795 | 0.812 |
| $T = 3$ | 0.968 | 0.924 | 0.944 | 0.797 | **0.741** | 0.957 | 0.903 | 0.794 | 0.810 |
| $\alpha = 0.5, \beta = 0.5$ | **0.971** | 0.926 | **0.950** | **0.820** | **0.736** | **0.960** | **0.915** | **0.821** | **0.820** |
| $\alpha = 1, \beta = 3$ | 0.970 | 0.926 | 0.943 | 0.790 | 0.684 | 0.957 | 0.905 | 0.799 | 0.803 |
| $\alpha = 1, \beta = 5$ | 0.970 | **0.928** | 0.946 | 0.776 | 0.710 | 0.959 | 0.907 | 0.800 | 0.810 |
| $\alpha = 2, \beta = 2$ | 0.970 | 0.926 | 0.947 | 0.784 | 0.704 | 0.959 | 0.907 | 0.800 | 0.805 |
| $\alpha = 5, \beta = 1$ | 0.969 | 0.926 | 0.944 | 0.693 | 0.695 | 0.958 | 0.907 | 0.792 | 0.803 |

### 5.3 Ablation Studies on Segmentation Performance

We further evaluate the contribution of each key module adapting non-contrast segmentation in our model architecture with the internal testing cohort. We optimize each module's performance by adapting the variation of hyperparameters 1) temperature scaling T and 2) the random weighting distribution $(\alpha, \beta)$.

**Variation of hyperparameters $T$:** From Figure 6(a), interestingly, constant robustness in performance is demonstrated across the variability of T and the median Dice with T=0.1 is slightly higher than the others. However, the segmentation performance did not substantially vary and may be due to the limited data augmentation performed for non-contrast patches in each batch (A=1) with limited GPU memory.

**Variation of weighting distribution $(\alpha, \beta)$:** We further investigate the influence of the hyperparameter $h'$ and adapt random intensity augmentation with different weighting on the contrasting context in the training process. Figure 6(b) illustrates the effect on the segmentation performance of extracting $h'$ with multiple shape parameters combination of a beta distribution. The segmentation performance with beta-distribution of shape parameters alpha=0.5 and beta=0.5, demonstrates significantly better generalizability across the dataset compared with the extracted $h'$ from other beta-distribution. The model with such beta distribution adopts the contrasting domain knowledge by heavily randomizing the weighting to either one of the domains. The increase of sensitivity is demonstrated by encoding the variability across the domain shift and is beneficial in adapting the structural information of the abdominal organs.

## 6. Discussions

In this work, we present a novel 3D anatomical-aware semi-supervised learning scheme ContrastMix to adapt non-contrast imaging for robust abdominal organs segmentation. One of the challenges in non-contrast segmentation is distinguishing the subtle boundary between organ interests. Unlike traditional generative adversarial approaches, we use pairwise-registered contrast-enhanced imaging to generate refined boundary context and provide morphological constraints as additional supervision. First, we provide a 3D coarse-to-fine pipeline, which refines the morphological context in organ-aware regions. Next, we sharpen the probabilistic context of the non-contrast boundary through the self-predicted prior and additionally provide teacher context from contrast-enhanced domain as supervision. Finally, we randomly weight the organ-wise contrast correlation between domains and adapt the generalized contrast context for robust segmentation. Moreover, a large scale of experiments is performed, including the comparison between current learning state-of-the-arts and ablation studies on the proposed innovations. We demonstrate that ContrastMix outperforms all fully supervised approaches for multi-organ segmentation. Additionally, we deploy our trained model on another unseen dataset for single organ segmentation and demonstrate the generalizability across different cohorts. The ablation studies provide a better understanding of the impact on distilling self-predicted context and the random intensity mixing strategy for non-contrast segmentation.

Although ContrastMix tackles the current challenges of non-contrast segmentation, limitations still exist in the process of ContrastMix. One limitation is the dependency of the teacher prediction quality in the source domain. As we leverage the morphological context in contrast-enhanced domain as supervision, low-quality predictions of organ-aware regions may also be possible to compute and use as guidance. Inaccurate prior information may thus be introduced into the training process. Another limitation is performance in an organ-centric setting with well aligned anatomical references. We aim to innovate a single stage pipeline with end-to-end optimization as our future work.

ContrastMix contributes to adapting target domain information without generating fake context and using target ground-truth labels for segmentation purposes. From the application perspective, it can be further extended in other anatomical locations such as brain CT. *Patel et al.* proposed a fully supervised network to perform intracerebral hemorrhage segmentation using non-contrast brain CT [35]. *Kuang et al*. proposed an adversarial generative network for stroke lesion segmentation in non-contrast brain CT [36]. Apart from the imaging of different anatomical locations, ContrastMix provides an opportunity to adapt other pairwise imaging modalities such as MRI and PET. Further exploration can be investigated by using CT as a reference domain and adapting multiple modalities for segmentation tasks within a single network architecture. In the model architecture perspective, disentanglement learning can be further investigated with the basis of ContrastMix. Shape and contrast representation can be separately extracted by the corresponding encoder and minimizing the effect of the contrast variation for segmentation tasks. *Chartsias et al.* proposes a style and content disentangling method to perform cardiac segmentation [37]. *Shin et al.* proposes an unsupervised disentanglement learning method to adapt the tri-planar gradient image representation for non-contrast small bowel segmentation [38]. Furthermore, self-supervised approaches can be additionally performed to separate the organs' shape representation into independent embedding. *Lee et al.* propose a semantic-aware contrastive learning approach to define semantic embedding for boosting segmentation performance [39]. Separating contrast and shape representation within a single network may be a promising direction to adapt multi-modality imaging with fewer hyperparameters configurations.

## 7. Conclusion

In summary, the proposed ContrastMix network achieved consistent performance on organ segmentation with non-contrast scans, compared with the current state-of-the-art approaches. The core innovations of ContrastMix are to 1) sharpen the certainty of the non-contrast boundary context with knowledge distillation and 2) adapt the contrast and morphology variations by generating samples with randomly weighted contrast and prior knowledge across domains. As our proposed pipeline focuses on adapting the contrast variation across domains, a potential extension can be focused on adapting significant domain shift of other imaging modalities for organ segmentation in unsupervised setting.

**Acknowledgements**

This research is supported by the NIH Common Fund and National Institute of Diabetes, Digestive and Kidney Diseases U54DK120058, NSF CAREER 1452485, NIH 2R01EB006136, NIH 1R01EB017230, and NIH RO1NS09529. ImageVU and RD are supported by the VICTR CTSA award (ULTR000445 from NCATS/NIH). We gratefully acknowledge the support of NVIDIA Corporation with the donation of the Titan X Pascal GPU usage.